# Unconventional superconductivity in the Kondo-lattice system $CeCu_2Si_2$ - a personal perspective

Frank Steglich[1,2]

[1]Max Planck Institute for Chemical Physics of Solids, 01187 Dresden, Germany
[2]Center for Correlated Matter and School of Physics, Zhejiang University, Hangzhou 310058, China

In the first part of this article, I briefly review early research activities concerning strongly correlated electron systems, beginning with the discovery of superconductivity and the first observation of a resistance minimum in nominally pure Cu-metal, which was explained many years later by Kondo. I will also address the antagonistic behavior of conventional (BCS) superconductivity and magnetism. The main focus of this paper is on the discovery of non-BCS-type superconductivity in the Kondo-lattice system $CeCu_2Si_2$, a prototypical heavy-fermion metal. Here, the superconducting state is created by a periodic lattice of 100% of magnetic $Ce^{3+}$ ions. Meanwhile, more than fifty lanthanide-, actinide-, and transition-metal-based intermetallic compounds are known to belong to the class of heavy-fermion superconductors. Finally, I will give my personal view of the current knowledge of this kind of unconventional superconductivity.

## I. INTRODUCTION

Research on strongly correlated electron systems (SCES) is of central importance in contemporary condensed-matter physics [1-3]. Originally devoted to *f*-electron systems only, it is meanwhile dealing also with *d*-electron-, and even *p*-electron materials [4,5]. Topics of current interest are, among others, pertinent problems such as topological properties of correlated insulators [6-10] and (semi)metals [11-15] as well as phenomena which are potentially relevant for applications, like a giant thermoelectric power factor [16-20]. Another subject of current research is the so-called *strange-metal* behavior of correlated materials [21-27] which has been discussed in connection to astrophysics [28]. It is fair to say that both correlation-driven properties of *f*-electron materials and unconventional superconductivity have remained cornerstones of SCES research.

This article begins with a chronological survey of the very early stages and even prenatal periods of SCES research, from the initial studies of the Kondo effect and superconductivity to early reports on Kondo-lattice systems and heavy-fermion phenomena (Sec. II), leading to the discovery of the first unconventional superconductor CeCu$_2$Si$_2$ (Sec. III). In Section IV, I shall present my personal view of the current status of heavy-fermion superconductivity, with special emphasis on the latter material. This survey is concluded in Section V.

## II. KONDO EFFECT *vs* SUPERCONDUCTIVITY, KONDO LATTICE SYSTEMS, HEAVY FERMIONS

The first indication of an increase of the electrical resistivity upon cooling to below $T \approx 2$ K in nominally pure Cu was communicated in 1930 [29]. Subsequently, a low-$T$ minimum in the temperature dependence of the resistivity, $\rho(T)$, was established for Au metal [30]. By the beginning of the 1960s, it had become clear that such a resistivity minimum only occurs in the presence of isolated impurity spins [31]. This was explained in 1964 by Kondo [32] who studied the exchange interaction between the spins of the magnetic impurity and the conduction electrons in third-order perturbation theory; this holds above a limiting temperature, the Kondo temperature $T_K$.

Shortly after this breakthrough, Kondo's results could be verified by resistivity experiments on dilute Mo-Nb and Mo-Re alloys doped with Fe [33]. It should be noted that, when studying dilute alloys containing transition-metal impurities, it is commonly extremely difficult to extract the relevant single-impurity effects by avoiding inter-impurity correlations. Therefore, Triplett and Phillips [34] had to choose very low dopant concentrations, in the range of a few tens of ppm only, in order to determine the contribution of the single Fe and Cr impurities to the low-$T$ specific heat of the canonical Kondo alloys Cu$_{1-x}$Fe$_x$ and Cu$_{1-y}$Cr$_y$. In both cases, they were able to separate the impurity term from the raw data and, at sufficiently low temperatures, found it to be proportional to $T$, $\Delta C(T) = \gamma T$. Upon normalizing the coefficient $\gamma$ by one mole of the magnetic component, one obtains 1 J/K$^2$mol (for **Cu**Fe) and even 16 J/K$^2$mol (for **Cu**Cr). These values are very similar to those directly measured a few years later for the so-called Kondo-lattice systems, with 100 at% magnetic composition, see below. The results by Triplett and Phillips were considered striking evidence of a local Kondo resonance and are well described by the theory of a local Fermi liquid [35].

At the end of the 1960s, sufficiently pure rare earths had become commercially available, in particular thanks to K. A. Gschneidner, Jr., at Ames Laboratory [36]. Because the partially-filled 4$f$-shell (in rare earths) is more localized than the 3$d$-shell (in transition metals), single-impurity effects in lanthanide alloys can be studied up to an impurity concentration of $\approx$ 1at%, which exceeds the corresponding limit in transition-metal alloys by at least one order of magnitude. Dilute La-based alloy systems containing Ce-impurities, in particular (**La**,Ce)Al$_2$ [37-43] and (**La**,Ce)B$_6$ [44-47], have become model systems for such investigations.

A notable provisional end in the exploration of the single-ion Kondo effect was reached in 1975, when Wilson treated the Kondo effect by the renormalization-group approach [48]. This way, he could show that the Kondo-coupling strength strongly increases upon lowering the kinetic energy and that at $T = 0$, a Kondo singlet forms implying a bound state below the Fermi energy. Thus, the Kondo effect completely demagnetizes the partially filled 3$d$ (or 4$f$/5$f$) shell of the magnetic impurity in the zero-temperature limit. On the other hand, at sufficiently high temperature, the impurity spin is almost decoupled from the conduction-electron spins.

Superconductivity discovered in 1911 is manifested by a vanishing electrical resistivity [49] and the expulsion of magnetic flux [50]. According to Bardeen, Cooper and Schrieffer (BCS), this is caused by an instability of the Fermi liquid of conduction electrons mediated by electron-phonon coupling which results in the formation of isotropic ($s$-wave) Cooper pairs with total spin $S = 0$ [51]. Numerous studies over the years revealed superconductivity and magnetism to be antagonistic phenomena. This was exemplified in 1958 by Matthias and co-workers [52], who doped the superconductor La ($T_c \approx 5.7$ K) with 1at% of all the rare-earth elements and studied the evolution of the superconducting transition temperature upon the increasing count of 4$f$-electrons. They observed a distinct depression of $T_c$ as a function of the size of the rare-earth 4$f$-spin, rather than the effective 4$f$-moment, and concluded that the Cooper pairs are broken up by spin-exchange scattering of the conduction electrons from the magnetic impurity. This *paradigm concerning the interplay between superconductivity and magnetism* was elucidated by Abrikosov and Gor'kov in the framework of perturbation theory [53]. Their results were confirmed by, among others, M. B. Maple who studied the superconductor LaAl$_2$ ($T_c = 3.3$ K) with a small amount ($x$) of Gd substituted for La. He found his data to agree very well with the theoretical prediction in the whole range of Gd-concentration up to the critical concentration $x_c \approx 0.9$at%, at which the superconductivity is fully suppressed [54].

As noticed by Matthias et al. [52], the Kondo ion Ce$^{3+}$ has an especially strong effect on $T_c$, which was theoretically explained with the aid of a self-consistent treatment of the Kondo scattering of conduction electrons from magnetic impurities by Müller-Hartmann and Zittartz [55]. They showed that the pair breaking in a *Kondo superconductor* is temperature-dependent and becomes strongest near $T_K$ ($<< T_c$, the transition temperature of the host metal), which may cause *reentrant superconductivity*, i.e., its disappearance at low temperatures. Shortly after publication of the theoretical prediction [55], this was discovered for $(La_{1-y}Ce_y)Al_2$, $y \leq y_c \approx 0.6$at%, independently by groups at the University of Cologne and the University of California, San Diego [56-58]. For a certain part of the parameter space, the results of [55] predicted even *three* subsequent values of $T_c$, which was indeed observed by K. Winzer for the dilute alloy system $(La_{0.80}Y_{0.20})_{1-x}Ce_x$ [59]. At that time, I was a research assistant in Cologne at the department led by G. von Minnigerode and, since 1974, by D. K. Wohlleben.

As discussed throughout this article, the discovery of the superfluid phases of liquid $^3$He by Osheroff, Richardson and Lee in 1972 [60,61] has had a lasting impact on the development of SCES research. This is mainly based on the fact that, as shown by Leggett [62], the superfluidity of $^3$He invokes unconventional Cooper pairs of *p*-wave symmetry and total spin $S = 1$. In addition, Anderson and Brinkman [63] had proposed that the net attractive interaction between the $^3$He atoms is mediated by ferromagnetic paramagnons, see also [64].

In 1975, a strongly renormalized low-temperature Fermi-liquid phase was reported by Andres *et al.* for the paramagnetic hexagonal intermetallic compound $CeAl_3$ [65]. According to these authors, below 150 mK the measured specific heat is proportional to temperature, i.e., identical to the electronic contribution. The Sommerfeld coefficient $\gamma = 1.62$ J/K$^2$mol of $CeAl_3$ exceeds that of Cu metal by more than a factor of 2000, which signals a correspondingly large effective mass of fermionic quasiparticles, $m^*$. Likewise, a Fermi-liquid-type $T^2$ dependence of the electrical resistivity was observed [65]: $\Delta\rho(T) = AT^2$, where the huge $A$ coefficient is proportional to the square of the effective charge-carrier mass, $(m^*)^2$. The authors explained their observation by a strong resonance scattering of the conduction electrons from the trivalent Ce ions, forming a Friedel 4*f* virtual bound state at the Fermi level [66].

After learning about this discovery by Andres *et al.* [65], I became interested in the competition between the on-site Kondo effect and inter-site magnetic correlations and studied, along with a small group of students, the low-$T$ properties of the $(La_{1-x}Ce_x)Al_2$ system in the whole range of Ce-concentrations [67]. Fig. 1a displays the specific-heat results for $CeAl_2$, the antiferromagnetically ordered ($T_N = 3.9$ K) cubic sister compound of $CeAl_3$, down to $T = 20$ mK [68]. At the lowest temperatures, there exists a nuclear contribution, while the specific heat is dominated by a magnon-derived $T^3$ term and contains an enhanced electronic contribution, $\gamma T$, with $\gamma = 135$ mJ/K$^2$mol. This $\gamma$-value is more than 20 times larger than the Sommerfeld coefficient of the non-magnetic reference compound $LaAl_2$. Our results were analyzed [67] together with K. D. Schotte, a theorist from the Free University of Berlin, and C. D. Bredl, a student of mine, see below. By taking the afore - mentioned value of $\gamma$ measured in the antiferromagnetic phase of $CeAl_2$ as an experimental input parameter and by replacing its complicated magnetic structure [69] by a mean field, we could describe the specific heat of this *Kondo-lattice* system ($T_K \approx 5$ K) with the aid of the so-called resonance-level model proposed by Schotte and Schotte [70], which explains very well the thermodynamic properties of *isolated Kondo impurities*. After switching off the mean field we could follow, within this oversimplified analysis, the evolution of the Sommerfeld coefficient upon cooling in the putative paramagnetic phase at $B = 0$ (dashed line in Fig. 1b) [67]. For $T = 0$, we found $\gamma \approx 1.7$ J/K$^2$mol, very similar to the value directly measured for the paramagnetic compound $CeAl_3$ [65], with almost the same $T_K$ value as $CeAl_2$ [71,72]. This result was an *eye-opener*, strongly suggesting that the factor of a thousand enhancement of the effective charge-carrier mass in the low-$T$ phase of such Ce-based compounds should be ascribed to the many-body Kondo effect rather than a one-electron mechanism, such as the formation of a Friedel virtual bound state [66].

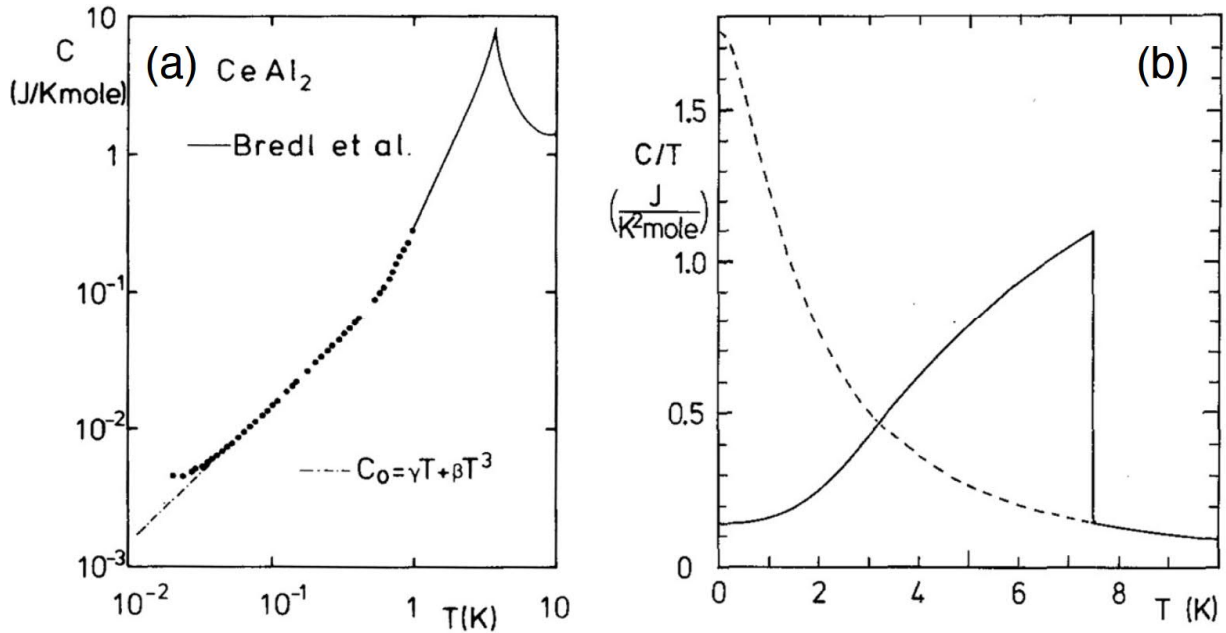

Fig. 1. (a) Molar specific heat of $CeAl_2$ vs temperature [68]. The solid line represents data from [67] for $T \geq 1$ K. (b) The solid line shows the electronic contribution as $C(T)/T$, calculated for $CeAl_2$ assuming antiferromagnetic order of mean-field type [67]. The dashed line displays the temperature dependence of $C(T)/T$ for single $S = 1/2$ Kondo ions with $T_K = 5$ K [70].

## III. CeCu$_2$Si$_2$: THE FIRST UNCONVENTIONAL SUPERCONDUCTOR

A relevant question raised by the report of a strongly renormalized Fermi-liquid phase in $CeAl_3$ [65] was: could such a *metallic* material possibly show

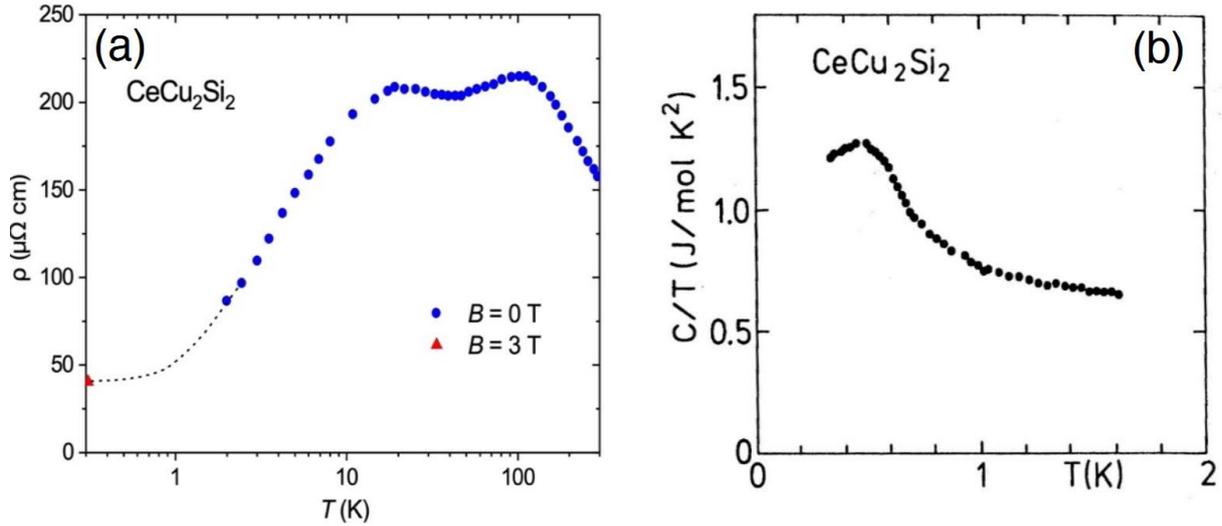

Fig. 2. Transport and thermodynamic properties of polycrystalline $CeCu_2Si_2$. (a) Electrical resistivity $\rho$ vs $T$ on a logarithmic scale [74], replotted from [73], and (b) specific heat as $C/T$ vs $T$ [75].

*unconventional* superconductivity, similar to the superfluidity in charge-neutral liquid $^3$He (with *p*-wave and spin-triplet Cooper pairs) which, in its normal state above $T_c \approx 1$ mK, also exhibits a strongly renormalized Fermi liquid phase [62,64]?

This came back to mind in 1977/78 when we, in a collaboration with D. K. Wohlleben's group, detected quite accidently superconductivity in the tetragonal compound $CeCu_2Si_2$. On studying the transport properties of this material [73], we were surprised to find that at the lowest accessible temperature of about 1.5 K, the electrical resistivity still showed a pronounced temperature dependence, with no tendency of saturation (Fig. 2a). To determine the residual resistivity, the sample was mounted into my home-made $^3$He cryostat, equipped with a superconducting magnet, and found to lose its resistivity near half a Kelvin. A magnetic field as high as 3 T had to be applied to suppress this unusual superconductivity. Simultaneously, the specific-heat coefficient $C(T)/T$ was determined by C. D. Bredl on another sample from the same batch [75]. He observed huge values and a hump in $C(T)/T$ at about 0.5 K (Fig. 2b), which the authors of [73] tentatively ascribed to magnetic correlations. Because the x-ray patterns of these argon-arc-melted polycrystals displayed a number of strange reflections, it was concluded that the superconductivity was most likely not intrinsic, but due to some unidentified spurious phase [73]. I was not fully convinced of this view for two reasons: first, the hump in $C(T)/T$ could alternatively indicate a (broadened) 2$^{nd}$ - order phase transition, and second, there was the exciting possibility of a metallic analogue to the superfluidity in liquid $^3$He. Therefore, I found it necessary to repeat the measurements on substantially purer samples.

It was fortunate that a few months after publication of this paper I met H. Schäfer, the director of the Eduard Zintl Institute for Inorganic Chemistry at the Technical University (TH) of Darmstadt, where I started a professorship in the fall of 1978. He was an expert in intermetallic compounds with the tetragonal $ThCr_2Si_2$ crystal structure, and I told him about our problems with $CeCu_2Si_2$ specimens lacking phase purity. I asked him to prepare samples of improved quality for us, which were provided pretty soon and subsequently studied via transport and thermodynamic measurements in our Cologne laboratory. My collaborators working there were Jan Aarts, a student of P. F. de Châtel's at the University of Amsterdam, who did part of his dissertation work in our lab, the doctoral students Claus-Dieter Bredl, Wolfgang Franz and Winfried Lieke as well as Dieter Meschede, who was about to finish his Diploma thesis. By February 1979, sufficient data on these new samples had been obtained, which convincingly showed that $CeCu_2Si_2$ indeed exhibits *bulk superconductivity* at $T_c \approx 0.5$ K [76].

As seen in Fig. 3a (main part), the low-$T$ electrical resistivity of normal-state $CeCu_2Si_2$ is linear in temperature which, in retrospect, indicates *non-Fermi- liquid* or *strange-metal* behavior. At $T_c$, the ac-susceptibility undergoes a sharp change from a huge Pauli spin susceptibility [77] to a pronounced diamagnetic value (inset of Fig. 3a). Fig. 3b displays a giant jump of the specific-heat coefficient, which is of the same order ($\approx 1$ J/K$^2$mol) as that of the $\gamma$-coefficient in the normal state when extrapolated to $T = 0$. This exceeds the Sommerfeld coefficient of the electronic specific heat in a simple metal by more than three orders of magnitude and indicates that the specific heat as measured in this low-temperature range is, like in $CeAl_3$ [65], identical with the electronic contribution.

Due to the *lattice* Kondo effect [78], in both $CeCu_2Si_2$ and $CeAl_3$ [71] slowly propagating Kondo singlets (i.e., super-heavy charge carriers, called *heavy fermions* [76]) are formed which in $CeCu_2Si_2$, as derived from the huge jump height in $C(T)/T$ at $T_c$ (see Fig. 3b), build up massive Cooper pairs with a short coherence length [79]. This was corroborated by a giant initial slope of the upper critical field curve at $T_c$, $B_{c2}' = -dB_{c2}/dT \approx 17$ T/K [79]. Note that, if the Cooper pairs were formed by the (co-existing) light conduction electrons, the resulting jump of the electronic-specific-heat coefficient would have been much too small to be resolved from the scatter of the data. Unannealed samples of the type-II superconductor $CeCu_2Si_2$ [79] showed a Meissner volume of about 2% only, apparently because the flux expulsion was strongly hampered by pinning centers. However, we were able to achieve an efficient reduction of the concentration of pinning centers through powdering and reannealing the samples, a procedure which greatly enhanced the Meissner volume to about 60% [79].

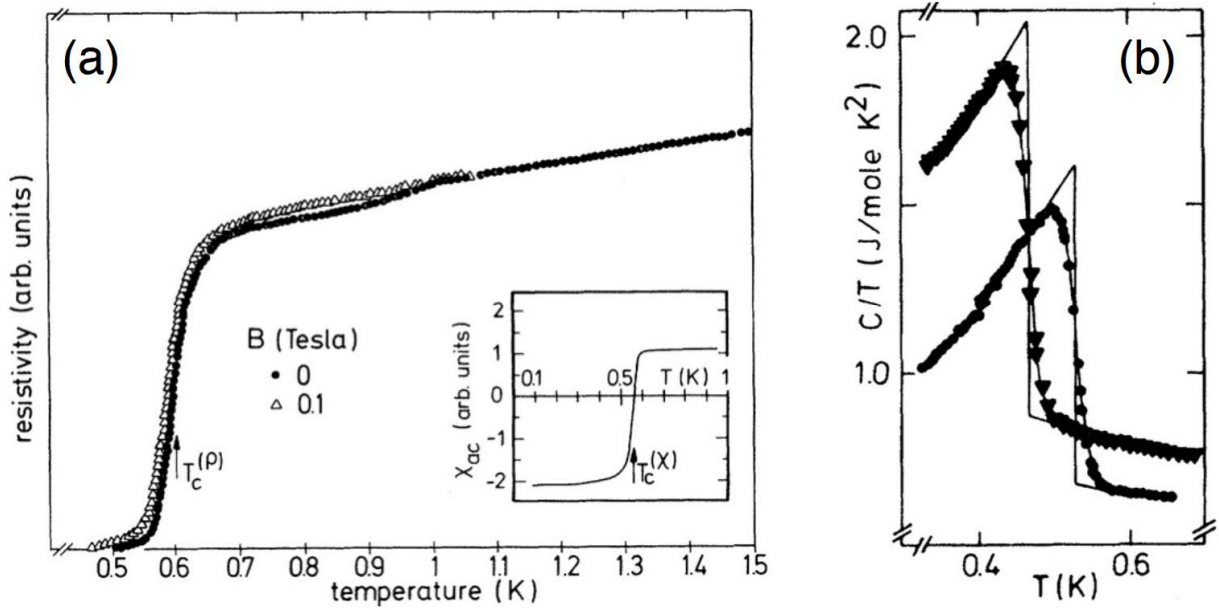

Fig. 3. (a) Resistivity $\rho(T)$ and ac-susceptibility $\chi_{ac}(T)$ as well as (b) specific heat as $C/T$ vs $T$ for polycrystalline CeCu$_2$Si$_2$, demonstrating bulk superconductivity at $T_c \approx 0.5$ K [76]. Note that the normal-state values of both $\rho(T)$ and $C(T)/T$ indicate non-Fermi-liquid behavior. In (b) results are shown for two different samples, with the same nominal composition and prepared in the same way, see text.

As the renormalized kinetic energy of the heavy quasiparticles, $k_B T_F^*$, is of the same order as the binding energy of the Kondo singlet, $k_B T_K$, [78] one obtains for CeCu$_2$Si$_2$, with $T_K \approx 15$ K [80], a ratio $T_c/T_F^* \approx 0.04$, compared to $T_c/T_F \approx 10^{-3}$–$10^{-4}$ for classical BCS superconductors. This highlights CeCu$_2$Si$_2$ as a *high-$T_c$ superconductor in a normalized sense* [76]. On the other hand, the ratio $T_K/\theta_D$, where $\theta_D$ is the Debye temperature, is only about 0.05, i.e., three orders of magnitude smaller than the ratio $T_F/\theta_D$ for main-group metals. This showed already at the outset [76] that the superconductivity in CeCu$_2$Si$_2$ cannot be explained by the weak-coupling BCS theory [51]. For, the latter relies on the retardation of the electron-phonon coupling by which the Coulomb repulsion between conduction electrons is minimized and, as a consequence, isotropic $s$-wave Cooper pairs with total spin $S = 0$ can be formed onsite. Compared to the Fermi velocity $v_F$ in a conventional superconductor, its renormalized counterpart $v_F^*$ in CeCu$_2$Si$_2$ is much smaller, at best of the size of the velocity of sound. Therefore, the electron-phonon coupling is not retarded which prevents the suppression of the Coulomb repulsion as well as the BCS-type onsite formation of Cooper pairs. In hind-sight, a magnetic pairing mechanism for CeCu$_2$Si$_2$ could have already been inferred from the observation that the quasiparticle entropy at $T_c$ is as large as a few percent of $R\ln 2$, the Zeeman entropy associated with the lowest-lying Kramers doublet of the crystal-field split $J = 5/2$ Hund's rule ground state of Ce$^{3+}$ [81-83].

In contrast to a BCS superconductor for which superconductivity is destroyed by a tiny amount of magnetic impurities, in $CeCu_2Si_2$ a periodic lattice of 100 at% of magnetic $Ce^{3+}$ ions appear to be necessary to generate superconductivity; for, the non-magnetic reference compound $LaCu_2Si_2$ lacking 4*f*-electrons is not a superconductor [76]. This unconventional nature of the superconductivity was confirmed by subsequent substitution experiments which revealed that the superconducting state in $CeCu_2Si_2$ is fully destroyed upon doping with a low concentration (1 at% or even less) of certain *nonmagnetic* impurities [84], which are harmless to conventional BCS superconductors [85,86], see Fig. 4a.

At the time, our report of bulk heavy-fermion superconductivity in $CeCu_2Si_2$ [76] was not fully acknowledged by the community, mainly because this apparently violated the afore-mentioned *superconductivity-magnetism paradigm*, but in part also because of weird *sample dependences* - like the ones displayed in Fig. 3b. Here, the results are shown for two specimens, which had been prepared and heat treated in very much the same way, but nevertheless exhibited quite different specific-heat values. These enigmatic sample dependences were not understood and could be explained only many years later with the existence of a *quantum critical point* (QCP) [87,88], see below. Because of this QCP, minor changes in the stoichiometry can lead to drastic changes of the physical properties o*f homogeneous* $CeCu_2Si_2$ samples: Stoichiometric (*A/S*-type) samples exhibit antiferromagnetic order below $T_N \approx 0.7$ K and superconductivity below $T_c \approx 0.55$ K, but at lower temperatures antiferromagnetic order becomes fully suppressed by the superconductivity. Samples with a tiny Cu-excess show superconductivity only, with $T_c \approx 0.6$ K (*S*-type). Less than a 1at% Cu-deficit results in *A*-type samples, which exclusively exhibit antiferromagnetic (spin-density-wave, SDW [89]) order below $T_N \approx 0.8$ K [90,91].

The reservations in the community against our discovery of bulk heavy-fermion superconductivity in $CeCu_2Si_2$ were finally overcome after a couple of years when excellent single crystals grown by W. Assmus' group [92] showed even more pronounced superconducting phase-transition anomalies than the polycrystals. Compared to the latter, these single crystals exhibited an even larger value of the initial slope of the upper critical field curve at $T_c$, i.e., $B_{c2}' \approx 23$ T/K (Fig. 4b), strongly supporting the notion that here, the Cooper pairs are formed by extremely heavy quasiparticles. The emergence of superconductivity in $CeCu_2Si_2$ had since been confirmed by several groups [93-96].

In 1983, H. R. Ott and H. Rudigier from ETH Zürich in collaboration with Z. Fisk and J. L. Smith from Los Alamos National Laboratory revisited $UBe_{13}$ [97], for which evidence of superconductivity had already been communicated earlier [98]. However, the authors of [98] concluded from the observation of an upper critical magnetic field which was unusually high in comparison to the low $T_c$ that the superconductivity was presumably not intrinsic, but rather due to filaments of elemental U precipitations. Studying flux-grown $UBe_{13}$ single crystals of improved quality, Ott et al. [97] were able to observe bulk heavy-fermion superconductivity with a huge specific-heat jump at $T_c \approx 0.85$ K, proving that this phenomenon is a general one, not restricted to a single material. The next heavy-fermion superconductor to be discovered in 1984 by G. R. Stewart and his colleagues at Los Alamos was $UPt_3$ [99]. In the same year, two additional U-based members of this family of new superconductors, $U_2PtC_2$ and $URu_2Si_2$, were reported by the Los Alamos group [100] resp. a Cologne/Darmstadt collaboration [101,102], see also [103,104].

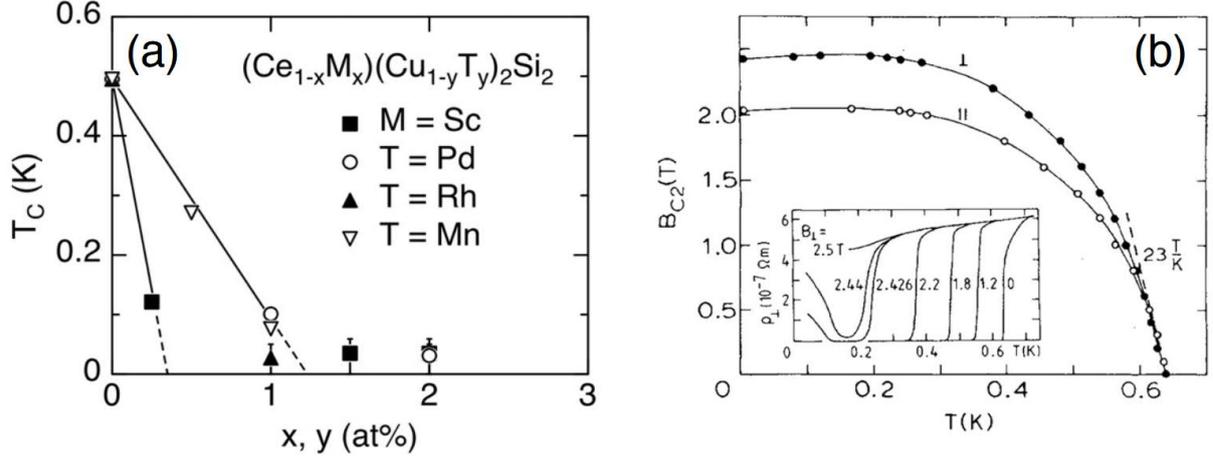

Fig. 4. (a) Dependence of the superconducting transition temperature of polycrystalline $CeCu_2Si_2$ on substituting nonmagnetic (and magnetic) dopants for Ce and Cu [3], replotted from [84]. (b) Upper critical magnetic field $B_{c2}$ vs $T$ of a $CeCu_2Si_2$ single crystal for fields applied within ($\parallel$) and perpendicular to ($\perp$) the Ce planes, as obtained from $\rho(T)$ measured parallel to the respective field [92]. Inset: $\rho(T)$ for different values of $B_\perp$. Note the re-entrant $\rho(T)$ behavior for $B_\perp \geq 2.4$ T, reflecting the shallow maximum of $B_{c2}(T)$ near $T \approx 0.2$ K (main part).

## IV. ON THE CURRENT STATUS OF HEAVY - FERMION SUPER-CONDUCTIVITY

The discovery of superconductivity in $CeCu_2Si_2$ and the advent of heavy-fermion physics in the early 1980s generated broad interest in unconventional superconductivity and strongly correlated electron systems. By 1992, the year of the first International SCES Conference, organized by T. Kasuya in Sendai, this research field had already substantially broadened [105,106]: new theoretical concepts had been invented to explain the physics of correlated materials, such as the slave-boson method [107,108] and dynamical mean-field theory [109,110].

For the first time, the heavy quasiparticles had been directly observed via magnetic quantum oscillation measurements [111] and in addition, there was great interest in exploring *Kondo insulators* [112] and *low-carrier-density Kondo systems* [113]. Three more heavy-fermion superconductors, i.e., $UNi_2Al_3$ [114] and $UPd_2Al_3$ [115] as well as $CeCu_2Ge_2$ under pressure [116], had been found at TH Darmstadt and the University of Geneva, respectively, increasing the number of known such superconductors to eight.

This figure has meanwhile increased to more than fifty [3,117,118], among them many Ce- and a few Pu-based tetragonal, so-called 115-materials as discovered by the Los Alamos group [119-122]. The Ce-based variants are prepared by increasing the *c/a* ratio of the cubic heavy-fermion superconductor $CeIn_3$ [123,124] upon inserting an additional layer of $TMIn_2$ (TM: Co, Rh or Ir). Related to the 115-materials, there also exist the 218-compounds [125-127] and their 127-counterparts [128]. Among all heavy-fermion superconductors, $PuCoGa_5$ exhibits a record-high $T_c = 18.5$ K [129]. Enhanced $T_c$ values are also observed for $PuRhGa_5$, $T_c = 8.7$ K [130], and $NpPd_5Al_2$, $T_c = 4.9$ K [131]. Further on, several compounds lacking an inversion-symmetry center have been added to this group of superconductors [132,133], following the discovery of heavy-fermion superconductivity in non-centrosymmetric $CePt_3Si$ in 2004 by E. Bauer and collaborators [134]. The lack of an inversion center may result in a mixing of even- and odd-parity pair states [135]. Meanwhile, heavy-fermion superconductivity has also been reported for an Fe-based intermetallic, $YFe_2Ge_2$ [136,137].

The physics of heavy-fermion metals is well described in terms of a competition between the Kondo effect, which tends to quench the local moments, and the Ruderman-Kittel-Kasuya-Yosida (RKKY) interaction, an indirect exchange interaction between the local magnetic moments mediated by the conduction electrons, by which the magnetic moments become stabilized [138-140]. These competing interactions give rise to a variety of different ground-state properties which cover Fermi-liquid and non-Fermi-liquid, superconducting and magnetically ordered phases [1,78,141-152]. The superconductivity in heavy-fermion compounds frequently occurs either coexisting with [114,115,134,153-155] or in the vicinity of [80,87,121] antiferromagnetic order. In several of these systems, spin-fluctuation-driven superconductivity [156] is found close to a QCP [80,123,124], at which antiferromagnetic order smoothly disappears. This is a consequence of the large amount of residual entropy that tends to be accumulated at the QCP, but is commonly eliminated by the formation of a novel symmetry-broken phase [157-163], preferentially unconventional superconductivity [164-168]. Interestingly, among the lanthanide-based heavy-fermion metals no superconductivity could yet be observed at a ferromagnetic QCP [22,169].

Various types of antiferromagnetic instabilities have been studied intensively over the past decades, *itinerant* spin-density-wave (SDW) [170-172] and valence-instability [173,174] QCPs as well as a *local* Kondo-destroying [175,176] one, see also [177]. The latter variant is a continuous zero-temperature partial Mott transition, an insight that was first derived by H. von Löhneysen and his collaborators from the discovery of unique *local* spin correlations [178] as seen in the inelastic neutron-scattering spectra of the quantum critical heavy-fermion metal $CeCu_{6-x}Au_x$ [179]. For the pressurized heavy-fermion superconductor $CeRhIn_5$, magnetic quantum-oscillation results demonstrated that its antiferromagnetic QCP coincides with an abrupt change of the Fermi volume, indicating that here, the magnetic instability is most likely of the Kondo-destroying variety [155,180]. Such an unconventional *local* type of QCP close to a superconducting phase was also proposed for both *β*-$YbAlB_4$ [181] and $YbRh_2Si_2$ [74,182,183], see below. Superconductivity driven by the critical fluctuations near a Kondo-destroying QCP has been theoretically explored by Hu et al. [184]. The quantum-critical phenomena observed for the two afore-mentioned Yb-based heavy-fermion metals as well as $CeRhIn_5$ have alternatively been ascribed to critical valence fluctuations [174].

In the vicinity of a heavy-fermion SDW QCP, at which the composite quasiparticles stay intact, a novel crossover scale $E^* = k_BT^*$ is anticipated to exist [160]. This scale, which is much smaller than the pure Kondo energy $k_BT_K$, is absent in ordinary transition-metal systems [185]. The quantum critical behavior is expected to be of the SDW type below $T^*$, while the critical fluctuations of the Kondo effect, i.e., partial Mott physics, are assumed to be operating above $T^*$ [160]. For $CeCu_2Si_2$ ($T_K$ = 15 K), $T^*$ was found to be 1-2 K [3]. Most interestingly, and in accord with theoretical predictions [182], the Cooper pairs in $CeCu_2Si_2$ appear to be formed with the aid of higher-frequency ($\omega > k_BT^*/\hbar$) fluctuations of the local 4*f*-spins [3]. On the other hand, the long-wavelength quantum-critical SDW-type fluctuations ($\omega < k_BT^*/\hbar$) seem to be pair breaking [3]. So far, a (three-dimensional) heavy-fermion SDW QCP could be identified unambiguously only for $CeCu_2Si_2$ [87-89,186,187], but it likely exists for a larger number of heavy-fermion superconductors, such as the canonical pressure-induced superconductor $CePd_2Si_2$ [123,124] and, as recently discovered, $CeSb_2$ under pressure [188].

For a few of the heavy-fermion superconductors, i.e., $UPd_2Al_3$ [189], $CeCoIn_5$ [190-192], $CeCu_2Si_2$ [80] and $UBe_{13}$ [193], inelastic-neutron-scattering results have revealed a *spin resonance* inside the superconducting gap, 2*Δ*. Recently, for the putative triplet superconductor $UTe_2$ [194-197], a spin resonance was discovered to be situated above the gap energy, at about 4*Δ* [198,199]. For the first three materials, which show superconductivity either coexisting with ($UPd_2Al_3$) or in the vicinity of long-range antiferromagnetic order, the resonance occurs exactly at the antiferromagnetic propagation wave vector and is part of a dispersive excitation mode [200]. For UTe$_2$, the spin resonance is found at an incommensurate wave vector and, like for the three former systems, to exhibit an *upward* dispersion relation [198]. This signals the existence of an antiferromagnetic paramagnon, highlighting a sign-changing superconducting order parameter [201,202]. Further on, in heavy-fermion superconductors, the spin resonance emerges out of a quasielastic response in the normal state. This is not observed for high-$T_c$ cuprate superconductors [200] where, below the energy of the spin resonance, the latter shows a *downward* dispersion relation (*hour-glass* dispersion) and is ascribed to a singlet-triplet excitation of the condensate of *d*-wave Cooper pairs [203,204].

Magnetically-driven heavy-fermion superconductivity was proposed by the theorists very early. While Anderson [205] assumed spin-triplet, *p*-wave pairing similar to that in superfluid $^3$He [62,64], Miyake *et al*. [206] and Scalapino *et al*. [207] predicted the formation of *d*-wave pairs mediated by antiferromagnetic spin fluctuations, in line with the results of subsequent inelastic-neutron-scattering experiments on UPt$_3$ [153,154], UPd$_2$Al$_3$ [189] and CePt$_3$Si [208].

In UPd$_2$Al$_3$ with a $5f^3$ configuration, two more *localized* 5*f*-electrons appear to be responsible for the magnetic properties, implying a nearly atomic staggered moment of 0.85$\mu_B$ [209,210], and a stronger hybridized (*itinerant*) 5*f*-electron to be responsible for the heavy-fermion phenomena [189]. Due to this dual character of the 5*f*-electrons [211], heavy-fermion superconductivity ($T_c$ = 1.9 K) coexists microscopically with local-moment antiferromagnetic order ($T_N$ = 14.3 K) [189,209,210,212], see also [213]. A *magnetic exciton*, i.e., a dispersive crystal-field excitation of the localized $5f^2$ configuration, acts as the acoustic spin wave in the ordered phase. At the antiferromagnetic ordering wave vector, this acoustic magnon is softened by as much as 30% on going from the normal into the superconducting state, where it's position is about 1 meV [189] and agrees well with that of the dip and second peak structure in the tunneling spectrum of this material [214]. Similar to an optical phonon in a classical strong-coupling superconductor [215,216], this excitation of the localized 5*f*-electrons appears to act as *glue* for the superconductivity in UPd$_2$Al$_3$, which is carried by the itinerant 5*f*-electrons [189].

Heavy-fermion superconductors show an even greater variety of phase diagrams and gap structures than suggested by the afore-mentioned early theoretical works. Thus, for instance, two different superconducting phases in the presence of short-range antiferromagnetic order have been established for UPt$_3$ [153,154, 217-219], multifaceted behavior was observed for UBe$_{13}$ substituted by a low concentration of Th [220-222], and a *hidden-order* phase apparently exists in URu$_2$Si$_2$ [101-104,

223-226]. All of these last three materials exhibit a superconducting state with broken time-reversal symmetry [221,227-229]. For $CeRh_2As_2$, with a locally non-centrosymmetric crystal structure, two superconducting phases have been reported, suggesting even-parity pairing at low magnetic fields but odd-parity pairing at high fields [230]. On the other hand, the recently discovered heavy-fermion superconductivity in the high-pressure phase of $CeSb_2$ appears to be of even parity at low as well as high magnetic fields, exceeding the Pauli limit by more than one order of magnitude [188].

Two Yb-based heavy-fermion superconductors (with very low $T_c$) are known: *β-YbAlB$_4$* ($T_c \approx 80$ mK) shows intermediate-valence behavior [181], quantum criticality at ambient pressure [231] and a critical charge mode [232]. The canonical Kondo-lattice system $YbRh_2Si_2$ [233,234] exhibits weak anti-ferromagnetic order below $T_N = 70$ mK [233,235] and a *local* QCP at a rather small critical magnetic field [236,237]. This is illustrated by thermally broadened jumps in the field–dependences of isothermal magneto-transport properties [238,239] and related anomalies in thermodynamic quantities [240]. Further on, a violation of the fundamental Wiedemann - Franz law was concluded from the observation that, on the approach of the QCP by sufficient cooling, the electronic Lorenz ratio (i.e., the ratio of the electrical over the electronic thermal resistivity) is reduced by about 10% [241,242]. Strange-metal behavior has been observed for $YbRh_2Si_2$ in both the spin [160,233,237] and charge [160,233,237,243,244] channel. Upon sufficiently cooling $YbRh_2Si_2$ at the critical magnetic field, the resistivity is found to be linear in $T$ down to 10 mK [160,237] and, as recently observed, even to 1 mK [245]. Interestingly, the temperature dependence of the Sommerfeld coefficient changes from a logarithmic divergence above $T \approx 300$ mK to a power-law divergence (with small critical exponent) at lower temperature [160,237]. This disparate behavior of thermodynamic and transport properties on approaching the QCP was tentatively ascribed to a *breakup of the composite heavy fermions* [237]. Further on, relatively strong Kondo-lattice correlations have to be built up upon cooling to well below $T_K \approx 30$ K [234] to ensure the onset of quantum criticality at lower $T$ [246]. At ultra-low temperature, nuclear antiferromagnetic order emerges [182,247]. The authors of [182] conclude that this nuclear order strongly competes with the primary 4*f*-electronic order, such that the latter disappears smoothly at a $B \approx 0$ Kondo-destroying QCP [182,183]. In contrast, it has been argued that the primary order persists down to the lowest temperatures [247]. *Bulk heavy-fermion superconductivity* in $YbRh_2Si_2$ is observed to form at $T_c \approx 2$ mK [182]. This was explained by the fact that the primary 4*f*-electronic order, which is detrimental to superconductivity [160,233, 237], has become sufficiently weak upon cooling to $T \approx 2$ mK due to the competing nuclear order [182,183]. Interestingly, magnetic [182] and resistive

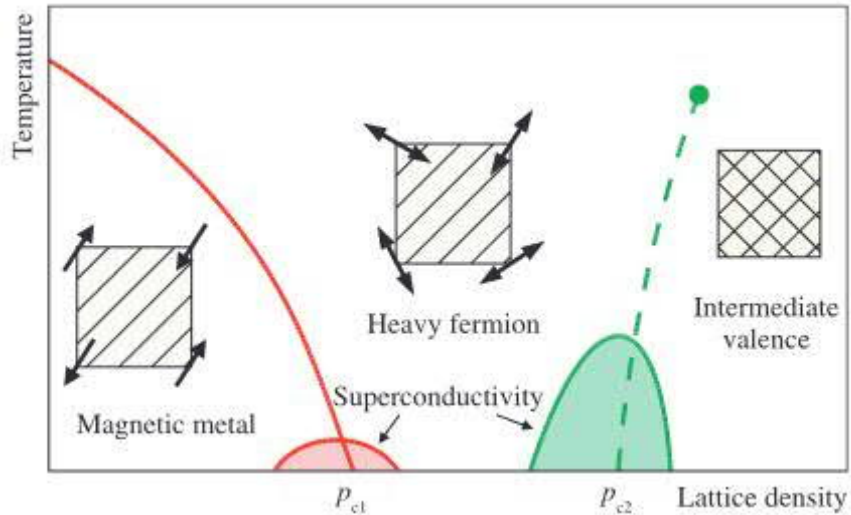

Fig. 5. Schematic phase diagram of heavy-fermion metals, like $CeCu_2Si_2$, as a function of pressure or lattice density. Two domes of superconductivity may exist, centered at $p_{c1}$, where antiferromagnetic order smoothly disappears, and at $p_{c2}$, at a putative valence-fluctuation-derived quantum critical point, i.e., where the critical point of a first-order valence-transition line is pushed to $T = 0$ by some additional tuning parameter [252].

[245] investigations reveal that *granular superconductivity* [183] sets in already at $T \leq 10$ mK.

$PrOs_4Sb_{12}$ shows a heavy-fermion normal-state and superconducting properties due to dominant quadrupolar (charge) rather than dipolar (spin) fluctuations [248]. In $PrM_2Al_{20}$ (M = Ir, Ti, V), superconductivity develops out of quadrupolar order [249-251]. Charge-fluctuation-driven superconductivity is also assumed to arise near a potential low-lying valence instability [173,174,252-254], eventually leading to a second superconducting dome as observed for $CeCu_2Si_2$, in which 10at% Si are substituted by Ge [252], see Fig. 5, and the Pu-based 115 materials [254].

While the majority of heavy-fermion superconductors are believed to show anisotropic even-parity Cooper pairing, a few of them are prime candidates for odd-parity pairing, i.e., $UPt_3$ [217,255], $UNi_2Al_3$ [256] and as already mentioned, $CeRh_2As_2$ at high magnetic fields [230]. Further members of this group are these distinct ferromagnets [257]: $UGe_2$ [258], $URhGe$ [259] and $UCoGe$ [260]. For the latter system, dominant longitudinal ferromagnetic spin fluctuations have been detected through angle-resolved NMR and Meissner experiments [261]. As mentioned before, $UTe_2$ is also believed to be a candidate for odd-parity pairing [194-197]. In addition, $UTe_2$ was proposed to be a *chiral topological superconductor* [262].

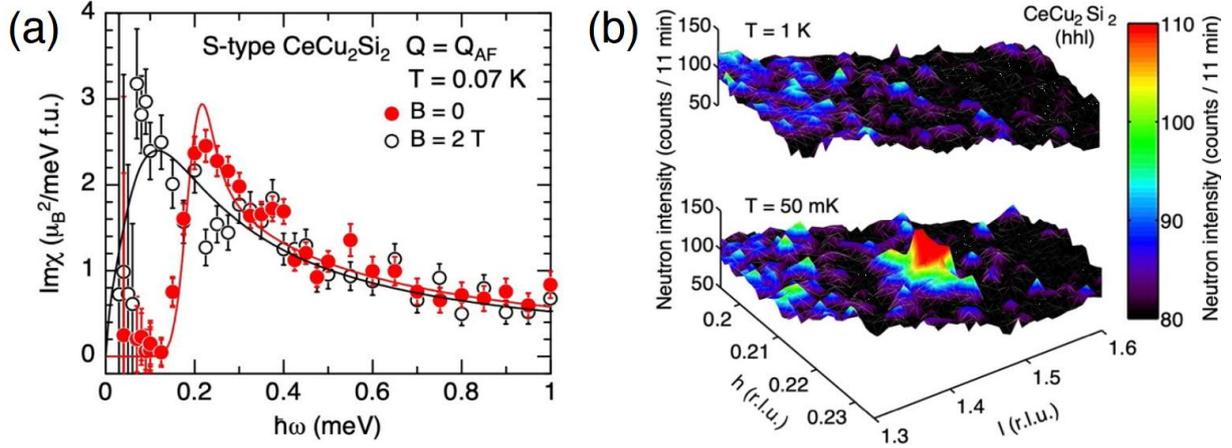

Fig. 6. (a) Low-energy spin excitations in a single crystal of *S*-type $CeCu_2Si_2$ at the SDW-ordering wave vector $Q_{AF}$ and $T = 0.07$ K in the superconducting ($B = 0$) and the normal state ($B = 2$ T) [80]. (b) Neutron-diffraction intensity map of the reciprocal (h h l) plane around the antiferromagnetic wave vector $Q_{AF} = (0.21\ 0.21\ 1.45)$ in a $CeCu_2Si_2$ single crystal of *A*-type at $T = 50$ mK and 1 K [89].

Exploring the physics of heavy-fermion superconductors often leads to the discovery of surprising, even unexpected physical properties [117,118,167, 222,257,263], as briefly exemplified in the following for the prototypical material $CeCu_2Si_2$ [3]. For many years, this compound was considered a model system for a one-band *d*-wave superconductor, with a nodal gap structure [264,265]. However, the results of precise low-temperature specific-heat measurements, reported in 2014 by the ISSP group at the University of Tokyo, revealed that the superconducting gap of $CeCu_2Si_2$ is completely open over the whole Fermi surface [266,267]. Subsequently confirmed by measurements of the penetration depth [268-270] and thermal conductivity [269] as well as new NQR experiments [271], this led to the notion that $CeCu_2Si_2$ is a fully-gapped two-band *d*-wave superconductor [3,270]. Alternative scenarios, describing isotropic (non-sign-changing) [268,269,272] and anisotropic [273,274] *s*-wave superconductors with $s_{++}$ - resp. $s_{+-}$ - Cooper-pair states, have also been proposed for this system.

The $s_{++}$ - scenario is highly unlikely by several reasons, in particular because of a pronounced peak in the inelastic neutron-scattering spectra of superconducting $CeCu_2Si_2$ (Fig. 6a) [80]. This peak is situated inside the superconducting gap, exactly at the propagation wave vector $Q_{AF}$ of the long-range SDW order (forming nearby in the phase diagram), as determined by neutron diffraction, Fig. 6b [89]. These findings highlight a sign-changing superconducting order parameter [201, 202], which is incompatible with $s_{++}$ - pairing.

Moreover, the antiferromagnetic ordering wave vector $Q_{AF}$ is identical with the nesting wave vector $\tau$ inside the dominating heavy-electron band as obtained by renormalized band calculations (Fig. 7(a)) [89, 147, 275,276] and indeed suppor-ted by recent ARPES results (Fig. 7(b)) [277] which indicates *intra-band* pairing. In contrast, for $s_{+-}$ - pairing, $Q_{AF}$ should be the same as the (*inter-band*) nesting wave vector connecting different electron and hole pockets at the Fermi surface [273,274]. Such an $s_{+-}$ - pairing is also highly unlikely for CeCu$_2$Si$_2$ because of the shape and effective mass of the hole pocket [3,277].

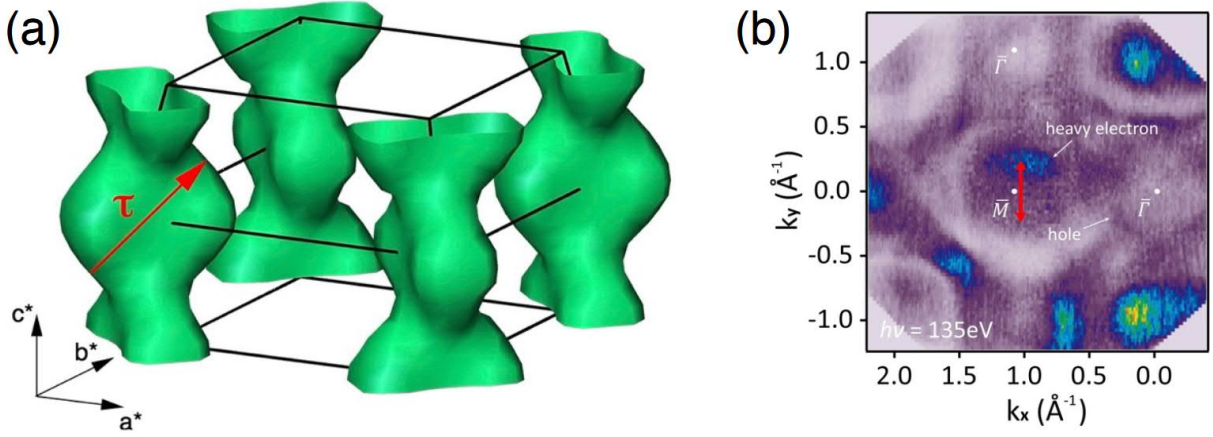

Fig. 7. Main heavy Fermi surface sheet in $CeCu_2Si_2$, indicating columnar nesting with wave vector $\tau$, (a) obtained by renormalized band theory [276] and found to be identical to the propagation wave vector $Q_{AF}$ from neutron diffraction [89]. (b) In-plane component of $\tau$ (red arrow) from ARPES experiments on an $S$-type single crystal [277].

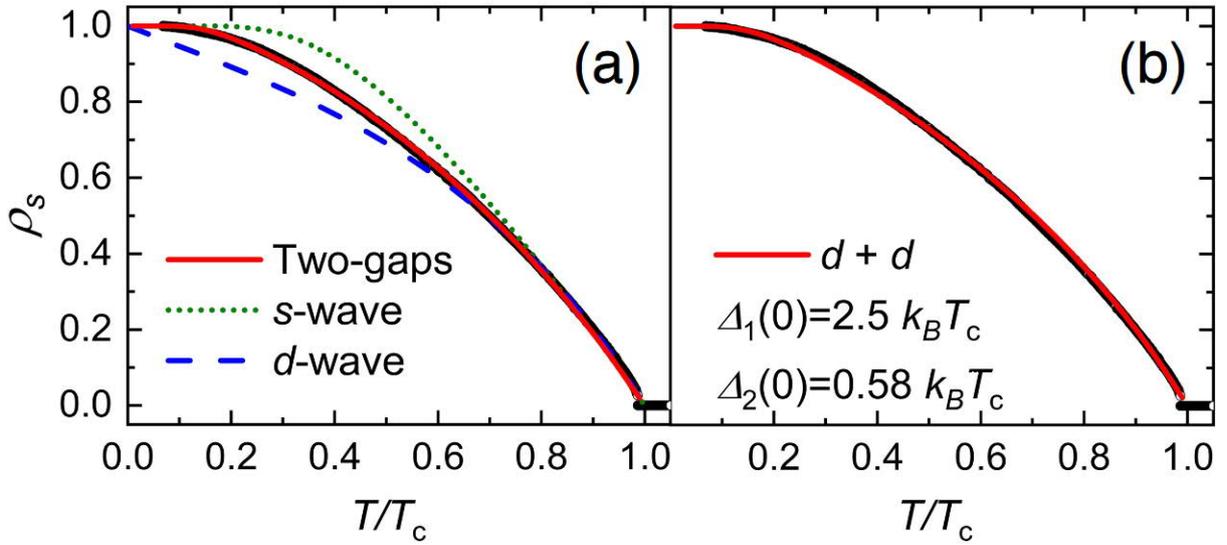

Fig. 8. Temperature dependence of the superfluid density derived from penetration-depth measurements on an $S$-type $CeCu_2Si_2$ single crystal, using the tunnel-diode-oscillator-based method [74], replotted from [270]. (a) Single-band $s$- or $d$-wave model cannot describe the data (black) below $T \approx 0.5T_c$. Two isotropic $s$-wave gaps (a large and a small one) yield a very good fit to the data over the whole range $T \leq T_c$. However, this model is unapt to explain the lack of the Hebel-Slichter peak at $T_c$ in the Cu-NQR data of [264,265]. (b) An excellent fit of the superfluid-density results (as well as of those for the specific heat [266,270] and Cu-NQR [264,265,271]) is achieved with a band-mixing ($d + d$) pairing model [270,278].

A band-mixing (*d+d*) pairing state [278] was shown to explain all presently available observations on CeCu$_2$Si$_2$ [3,270], including specific-heat and penetration-depth results (Fig. 8). The (*d+d*) pairing is very similar to what was proposed [279] for the Fe-based chalcogenide superconductors due to strong orbital-selective electronic correlations. It may be considered [278] a *d*-wave analogue to the spin-triplet pair states in the fully gapped *p*-wave phase, the B-phase, of superfluid $^3$He [62,64].

The afore-mentioned multi-orbital resp. multi-band character of the electronic correlations arguably allows for a larger variety of pair states compared to single-band superconductors [280]. This, most likely, enables the observation of a spin resonance in CeCu$_2$Si$_2$ [80] as well as UTe$_2$ [198,199]. For both compounds, distinct intra-band pairing components are anticipated to give rise to a *sign change* of the superconducting order parameter in the *fully-gapped* superconductor CeCu$_2$Si$_2$ [3,80,270,278] on the one hand and *antiferromagnetic spin fluctuations* in the candidate *triplet* superconductor UTe$_2$ [194-199] on the other [280].

# V. CONCLUDING REMARKS

To conclude this survey, I illustrate in Fig. 9 the evolution of those research fields devoted to the antagonistic phenomena *magnetism*, based on the repulsive Coulomb interaction and Hund's rule correlations, and *superconductivity*, originating in a net attractive interaction between fermionic quasiparticles [281]. Both fields have eventually merged, as first observed in 1972 when superfluidity, driven in particular by spin fluctuations at small momenta [64], was discovered in liquid $^3$He [60,61]. In 1979, this was found to be realized also for a metallic material, CeCu$_2$Si$_2$ [76], exhibiting superconductivity driven by antiferromagnetic spin fluctuations [80].

Electronically-driven Cooper pairings appear to operate not only in heavy-fermion superconductors [117,118] but also in organic charge-transfer salts [282-284], high-$T_c$ cuprate superconductors [2,4,285,286], Sr$_2$RuO$_4$ [287-289], Fe-based pnictides and chalcogenides [290-292], Moiré-structured materials [5,293,294] and, as discovered very recently, La$_3$Ni$_2$O$_7$ under high pressure [295,296], see also [164-168]. The entire topic of these unconventional *high-$T_c$ superconductors in a normalized sense* is likely to remain in the center of SCES research activities for many more years to come. This will be in parallel with an increasing interest in hydrogen-rich materials for which near-room-temperature superconductivity of the conventional variant, as indicated by a zero resistivity, was observed upon the application of extremely high pressure [297-300].

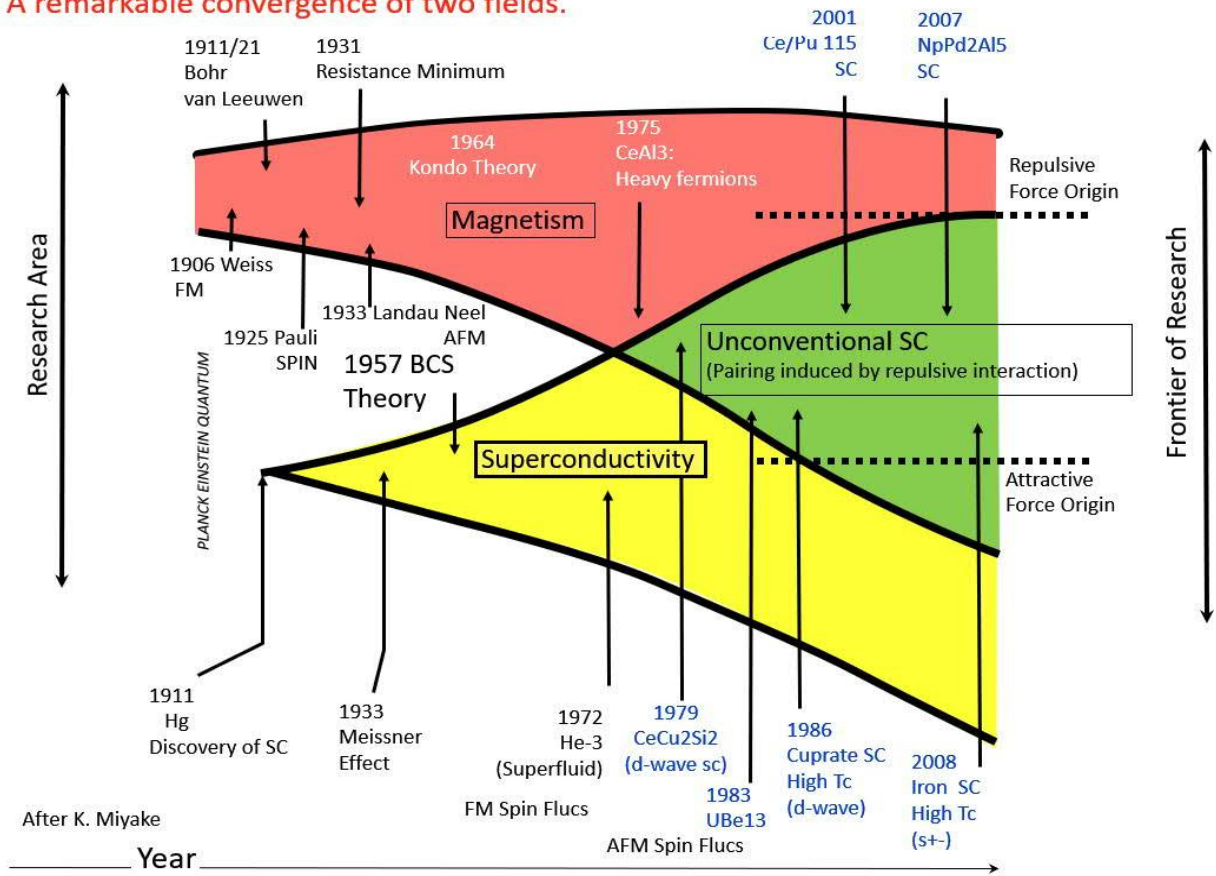

Fig. 9. The figure displays the evolution and merger of the research areas "magnetism" and "superconductivity", as described in the text [281].

## ACKNOWLEDGMENTS

I have benefited greatly from an in-depth discussion with Dieter Vollhardt and his critical comments. My sincere thanks go to Yang Liu, Kazumasa Miyake, Michael Smidman, Qimiao Si, Oliver Stockert, Chandra Varma and Steffen Wirth for a careful reading of the manuscript and several useful suggestions. I am grateful to Premala Chandra, Piers Coleman, Michael Lang, Jens Müller, Konrad Samwer and Huiqiu Yuan for stimulating conversations.